\documentclass{ws-procs9x6}            
\begin{document}
\title{Heavy $K^*(4307)$ meson with hidden charm}

\author{X.-L. Ren$^{1}$, B. B. Malabarba$^2$, L.-S. Geng$^3$, K. P. Khemchandani$^4$, and A. Mart\'inez Torres$^{2}$}

\address{$^{1}$Ruhr-Universit\"{a}t Bochum, Fakult\"{a}t f\"{u}r Physik und Astronomie, Institut f\"{u}r Theoretische Physik II, D-44780 Bochum, Germany\\
$^{2}$Instituto de F\'isica, Universidade de S\~ao Paulo, C.P. 66318, 05389-970 S\~ao 
Paulo, S\~ao Paulo, Brazil\\
$^3$School of Physics \& Beijing Key Laboratory of Advanced Nuclear Materials and Physics, Beihang University, Beijing 100191, China\\
$^4$Universidade Federal de S\~ao Paulo, C.P. 01302-907, S\~ao Paulo, Brazil}

\begin{abstract}
We report on a robust prediction of heavy $K^*$ meson by solving the  Faddeev equations with fixed-center approximation for the three-body $KD\bar{D}^*$ system.
As the excited Kaon state, $K^*$ is an exotic hidden charm meson with $M-i\Gamma/2=4307\pm2-i9\pm2$ MeV and $I(J^P)=1/2(1^-)$.
We further performed the evaluation of the decay width of $K^*(4307)$ to the open two-body channels. 
We expect that the above findings inspire an experimental investigation of this exotic $K^*$ meson and to study the so far unexplored heavy strange meson sector.
\end{abstract}


\bodymatter

\section{Introduction}\label{sec1} 
Recently, the exotic hadrons with the open/hidden heavy quark components have attracted great attention in the experimental and theoretical studies~\cite{Chen:2016spr,Chen:2016qju}.  
However, in the strange sector, there is no experimental data available on heavy $K$ or $K^*$ meson states, leaving the heavy strange physics experimentally unexplored. In this Hadron conference, the COMPASS collaboration presented their preliminary results of the strange-meson spectrum~\cite{Compass}.

From the theoretical side, we have explored the possibility of forming a heavy $K^*$ meson in the three-body $KD\bar{D}^*$ system~\cite{Ren:2018pcd}.
All the two-body interactions, $KD$, $K\bar{D}^*$, and $D\bar{D}^*$, are stringently constrained by a large number of experimental ($D_{s0}^*(2317)$, $D_{s1}^*(2460)$, $X(3872)$, and $Z_c(3900)$) as well as the lattice QCD data, which lead to a unique system of $KD\bar{D}^*$. Using the so-called fixed-center approximation (FCA)~\cite{Kamalov:2000iy} to the Faddeev equations, we found a heavy $K^*$ meson around $4307$ MeV with hidden charm. In order to provide more information, we also studied the decay width of $K^*(4307)$ from two-body decay processes~\cite{Ren:2019umd}.

\section{$K^*(4307)$ state in the $KD\bar{D}^*$ system}\label{sec2}

For a three-body system, if the mass of the third particle $P_3$ is much smaller than a stable cluster composed of the two other particles $P_1$ and $P_2$, one can 
 consider the use of the FCA to the Faddeev equations. Our $KD\bar{D}^*$ system exactly satisfies such criteria. We take the light kaon as the third particle to scatter off the cluster of $D\bar{D}^*$, which can form the as $X(3872)$ or $Z_c(3900)$ states~\cite{Gamermann:2006nm,Aceti:2014uea}.

Here, we briefly present the basic equations, in particular for the coupled-channel case, in the FCA framework. The details can be found in Ref.~\cite{Ren:2018pcd}.
The total scattering amplitude is decomposed into two Faddeev partitions, 
\begin{eqnarray}\label{Eq:FCA}
  T &=& T_{31} + T_{32},\nonumber\\
 T_{31} &=& t_{31} + t_{31} G_0 t_{32} + t_{31} G_0 t_{32} G_0 t_{31} + \cdots = t_{31} + t_{31} G_0 T_{32},\nonumber\\
 T_{32} &=& t_{32} + t_{32} G_0 t_{31} + t_{32} G_0 t_{31} G_0 t_{32} + \cdots = t_{32} + t_{32} G_0 T_{31},
\end{eqnarray}
where the two-body amplitudes, $t_{31}$ and $t_{32}$, are the functions of the isospin $0$ and $1$ $s$-wave interactions of the $KD$ and $K\bar{D}^*$ subsystems~\cite{Guo:2006fu,Guo:2006rp}.
\begin{figure}[b]
\centering
\includegraphics[width=0.46\textwidth]{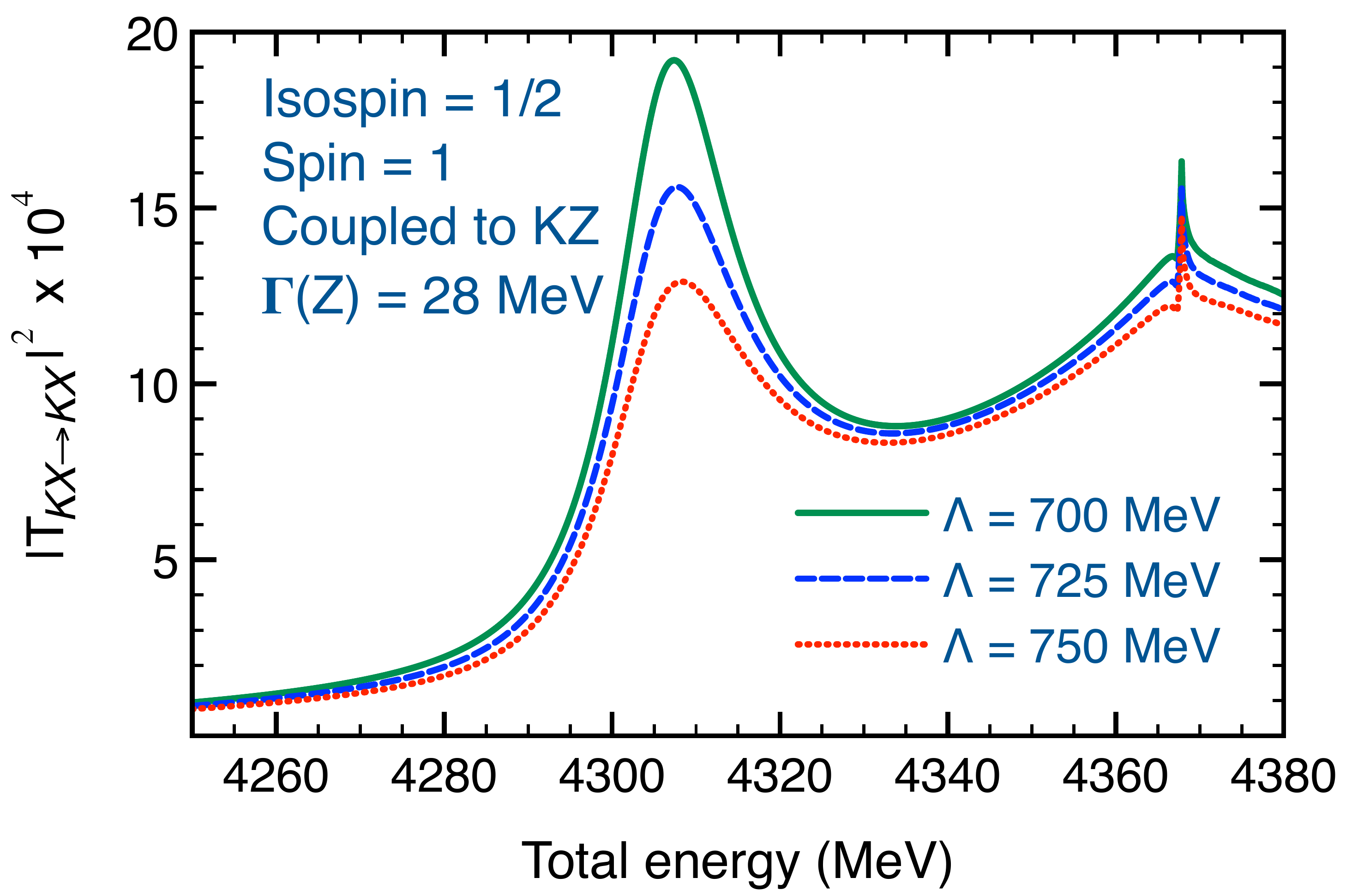}
\includegraphics[width=0.46\textwidth]{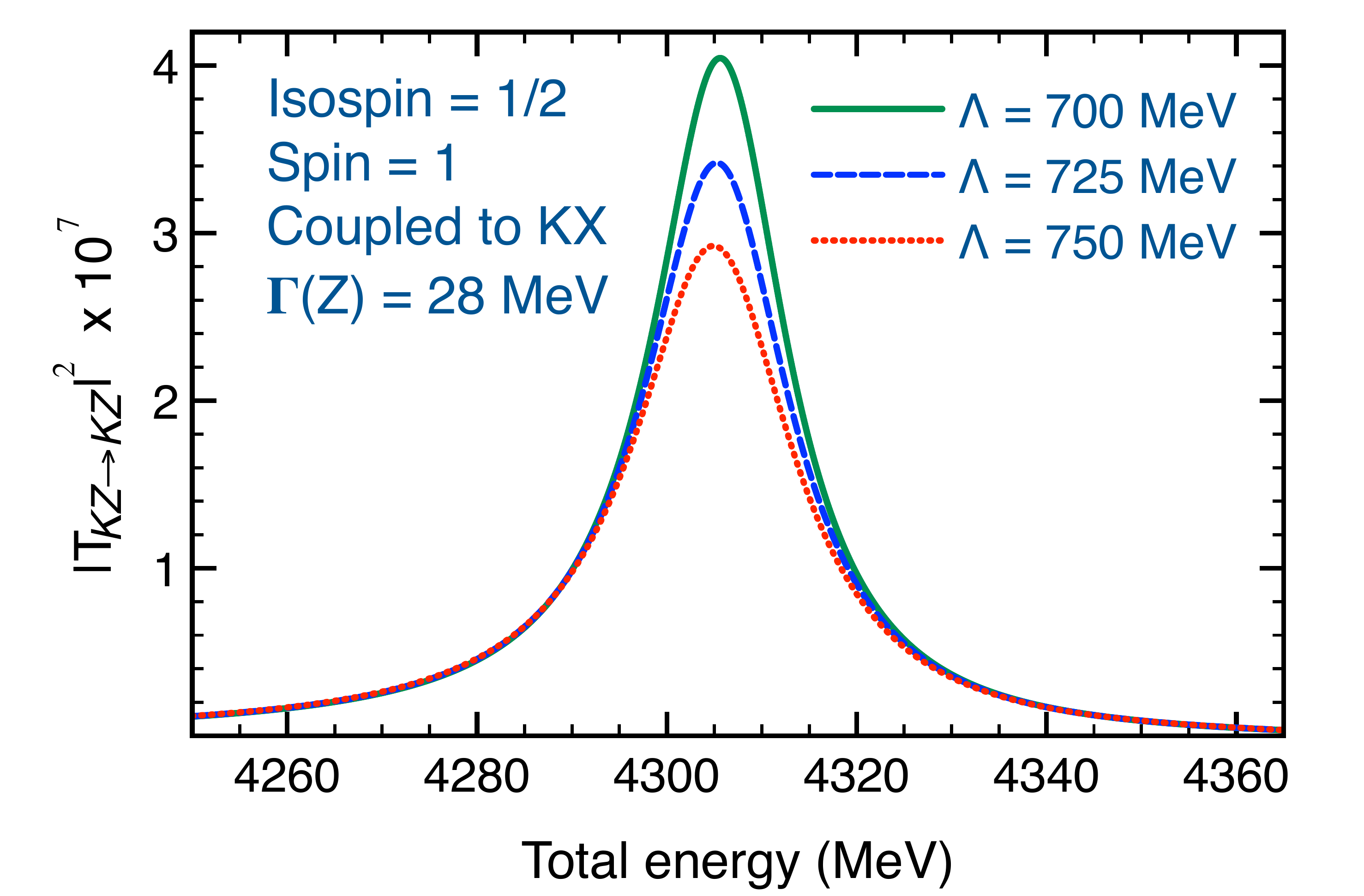}
\caption{Modulus squared of the $KX$ and $KZ$ scattering amplitudes in $I=1/2$ with different cutoffs. A cusp related to the three-body $KD\bar D^*$ threshold is observed in the $KX\to KX$ amplitude.}
\label{Fig:KXKZ}
\end{figure}

In the $K[D\bar{D}^*]$ system,  we consider $KX(3872)$ and $KZ_c(3900)$ as coupled channels with total isospin $I=1/2$. As shown in Ref.~\cite{Ren:2018pcd}, this makes that the $t_{31}$ and $t_{32}$ amplitudes can be written as $2\times 2$ matrices in the coupled channel space, 
\begin{equation}
	t_{31}=\begin{pmatrix}
		(t_{31})_{11} & (t_{31})_{12} \\
		(t_{31})_{21} & (t_{31})_{22} 
	\end{pmatrix},\quad 
	t_{32}=\begin{pmatrix}
		(t_{32})_{11} & (t_{32})_{12} \\
		(t_{32})_{21} & (t_{32})_{22} 
	\end{pmatrix},	
\end{equation}
where the element $(11)$ denotes $KX\to KX$, the element $(12)$ $KX\to KZ_c$, and so on. 
The corresponding loop function $G_0$ is,
\begin{equation}
G_0 = \begin{pmatrix}
	(G_0)_{11} & 0 \\
	0  &  (G_0)_{22}
\end{pmatrix}.	
\end{equation}
Finally, the total $T$-matrix appearing in Eq.~(\ref{Eq:FCA}) becomes as,
\begin{equation}
T=T_{31} + T_{32} = \begin{pmatrix}
	T_{11} & T_{12} \\
	T_{21} & T_{22}
\end{pmatrix}.
\end{equation}

In Fig.~1, the modulus squared of the $KX$ and $KZ_c$ scattering amplitudes in $I=1/2$ are given with the momentum cutoff varying from 700 MeV to 750 MeV. 
We found that the mass and width of the heavy $K^*$ meson in the $KX$ configuration is $M-i\Gamma/2 = (4308\pm1)-i(8\pm1)$ MeV, and of the $KZ_c$ configuration is $M-i\Gamma/2 = (4306\pm1)-i(9\pm1)$ MeV. 
After averaging, the mass of the $K^*$ meson is $4307\pm2$ MeV with a width of $18\pm 4$ MeV. 
Note that our result is  consistent with the one of Ref.~\cite{Ma:2017ery}, where the Born-Oppenheimer approximation is applied to the $KD\bar{D}^*$ system by solving the Schr\"{o}dinger equation. 

\begin{figure}[b]
\centering
\includegraphics[width=\textwidth]{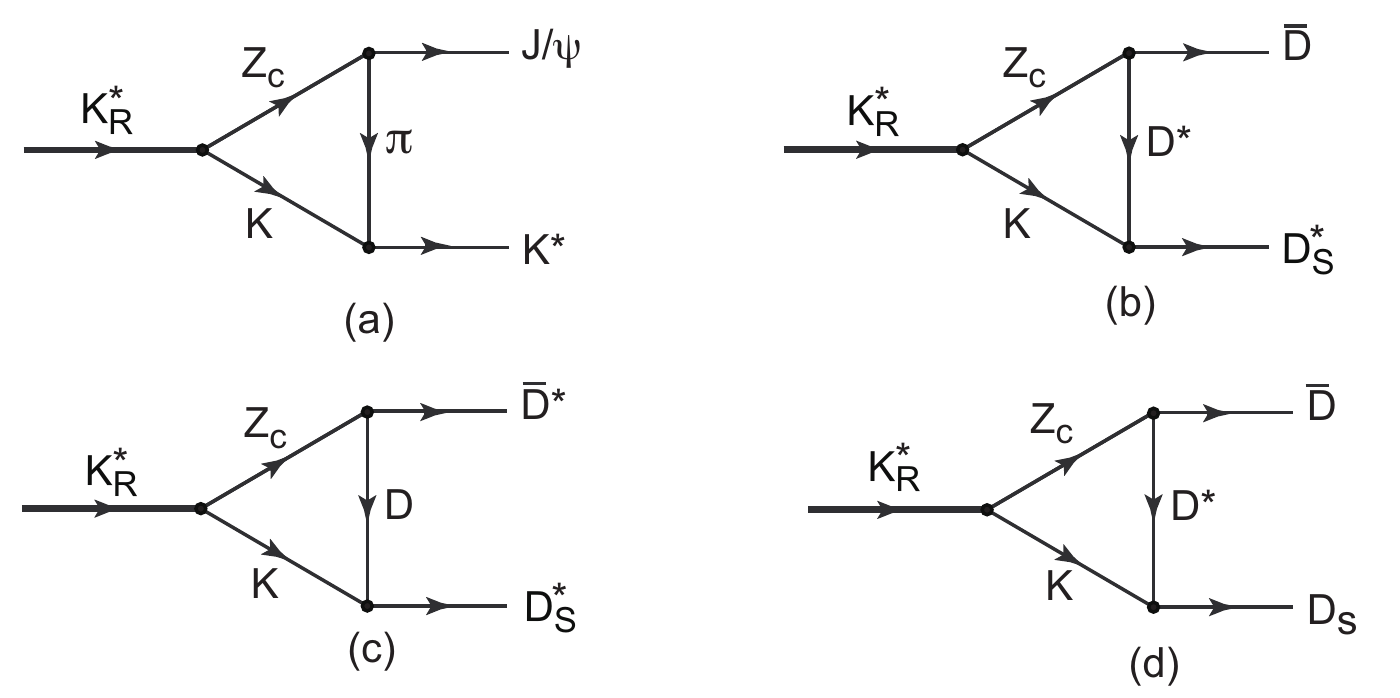}
\caption{Main two-body decay channels for the predicted $K^*(4307)$ meson.}
\label{Fig:decay}
\end{figure}

\section{$K^*(4307)$ decay}
\label{sec3}

In the following, we investigated the properties of the $K^*$ state and calculated the decay widths of the possible two-body channels. As shown in Fig.~1, the magnitude of the squared amplitude in the $KZ_c$ configuration is around 200 times larger than that found in the $KX$ configuration. 
Thus, the decay width of $K^*(4307)$ is mainly from the subprocess $K^*\rightarrow K Z_c$. In Fig.~2, the open two-body decay channels, $J/\psi K^*(892)$,~$\bar{D}D_s$,~$\bar{D}D_s^*$, and $\bar{D}^*D_s^*$, via the triangle loops are presented.

For the details about the determination of the $t_i$~($i=a,~b,~c,~d$) amplitudes and the calculation of the triangular loops using the momentum cutoff regularization, we refer the reader to Ref.~\cite{Ren:2019umd}.


Once we have the $t_i$, the decay width of the $K^*(4307)$ meson to the two-body channels can be obtained 
\begin{equation}
	\Gamma_{i} = \int\frac{d\Omega}{4\pi^2} \frac{1}{8 M_{K_R^*}^2} \frac{p_\mathrm{c.m.}}{3}\sum |t_i|^2 = \frac{p_\mathrm{c.m.}}{24\pi M_{K_R^*}^2}\sum |t_i|^2,
	\label{width}
\end{equation}
where $d\Omega$ represents the solid angle, $p_\text{c.m.}$ is the center of mass momentum of the particles in the final state, the factor $3$ has its origin in the average over the $K^*(4307)$ meson polarizations and the symbol $\sum$ indicates summation over the polarization of the initial and final states.

With the momentum cutoff 
$\Lambda$ changing from 700 MeV to 800 MeV, as used in Ref.~\cite{Ren:2018pcd}, the decay widths from different channels are
\begin{align}
\Gamma_{a}&=6.97\pm0.27~\text{MeV},\quad \Gamma_b=0.54\pm0.08~\text{MeV},\nonumber\\
\Gamma_c&=0.54\pm0.07~\text{MeV},\quad\Gamma_d=1.14\pm0.17~\text{MeV}.
\end{align}
Besides, we also considered the width of $K^*(892)$ in the evaluation of the decay width of $K^*(4307)\to J/\psi K^*$. Since the mass of $K^*(4307)$ is far from the $J/\psi K^*(892)$ threshold, the decay width $\Gamma_a$ does not change.

\section{Conclusion}
In this talk, we presented a prediction of heavy $K^*(4307)$ meson with hidden charm in the $KD\bar{D}^*$ system by solving the Faddeev equations with the fixed center approximation.
In order to provide more information about the internal structure on $K^*(4307)$, we calculated the decay width of $K^*(4307)$ to two-body channels. We are now working on the study of $B$ meson decay to $J/\psi \pi \pi K$, which could be handled by e.g. LHCb experiment to investigate the signal of $K^*(4307)$ meson in the $J/\psi \pi K$ invariant mass distribution~\cite{Ren:2019}. 
Furthermore, along this line, the possible bound states with hidden bottom are also predicted in the $\bar{K}^{(*)}B^{(*)}\bar{B}^{(*)}$ systems~\cite{Ren:2018qhr}. 
 We hope that the above studies could arouse interest to investigate the exotic states in the strange sector.

\section{Acknowledgement}
 X.-L.R. acknowledge the financial support from the Hadron 2019 conference. This work was partly supported by DFG and NSFC through funds provided to the Sino-German CRC 110 ``Symmetries and the Emergence of Structure in QCD'' (DFG Grant No. TRR110), the NSFC under Grants No. 11735003, No. 11522539, 11375024, and No. 11775099, Funda\c c\~ao de Amparo \`a pesquisa do estado de S\~ao Paulo (FAPESP) under Grants No. 2019/16924-3 and No. 2019/17149-3, and CNPq (Grant Nos. 310759/2016-1 and 311524/2016-8).
 

\begin{thebibliography}{10}

%
  
\bibitem{Chen:2016spr} 
  H.-X.~Chen, W.~Chen, X.~Liu, Y.-R.~Liu and S.-L.~Zhu,
  Rept.\ Prog.\ Phys.\  {\bf 80},  076201 (2017).
  
\bibitem{Chen:2016qju} 
  H.-X.~Chen, W.~Chen, X.~Liu and S.-L.~Zhu,
  Phys.\ Rept.\  {\bf 639}, 1 (2016).
  
\bibitem{Compass}
  COMPASS Collaboration [S. Wallner, et al.,], ``Strange-meson spectroscopy at COMPASS'' in the Hadron (2018) proceedings.
  
\bibitem{Ren:2018pcd} 
  X.-L.~Ren, B.~B.~Malabarba, L.-S.~Geng, K.~P.~Khemchandani and A.~Mart\'inez Torres,
  Phys.\ Lett.\ B {\bf 785}, 112 (2018).
  
\bibitem{Kamalov:2000iy} 
  S.~S.~Kamalov, E.~Oset and A.~Ramos,
  Nucl.\ Phys.\ A {\bf 690}, 494 (2001).
  
  

  
\bibitem{Ren:2019umd} 
  X.-L.~Ren, B.~B.~Malabarba, K.~P.~Khemchandani and A.~Mart\'inez Torres,
  JHEP {\bf 1905}, 103 (2019).
  
\bibitem{Gamermann:2006nm} 
  D.~Gamermann, E.~Oset, D.~Strottman and M.~J.~Vicente Vacas,
  Phys.\ Rev.\ D {\bf 76}, 074016 (2007).
  
  
\bibitem{Aceti:2014uea} 
  F.~Aceti, M.~Bayar, E.~Oset, A.~Mart\'inez Torres, K.~P.~Khemchandani, J.~M.~Dias, F.~S.~Navarra and M.~Nielsen,
  Phys.\ Rev.\ D {\bf 90}, 016003 (2014).
  
\bibitem{Guo:2006fu} 
  F.-K.~Guo, P.~N.-Shen, H.-C.~Chiang, R.-G.~Ping and B.-S.~Zou,
  Phys.\ Lett.\ B {\bf 641}, 278 (2006).
  
\bibitem{Guo:2006rp} 
  F.-K.~Guo, P.-N.~Shen and H.-C.~Chiang,
  Phys.\ Lett.\ B {\bf 647}, 133 (2007).
  
  
\bibitem{Ma:2017ery} 
  L.~Ma, Q.~Wang and U.-G.~Mei\ss ner,
  Chin.\ Phys.\ C {\bf 43},014102 (2019).
  \bibitem{Ren:2019} 
  X.-L.~Ren, K.~P.~Khemchandani and A.~Mart\'inez Torres, 
  arXiv:1912.03369 [hep-ph].
  
  
\bibitem{Ren:2018qhr} 
  X.-L.~Ren and Z.-F.~Sun,
  Phys.\ Rev.\ D {\bf 99}, 094041 (2019).

\end{thebibliography}



\end{document}